\documentclass[12pt,a4paper,reqno]{article}

\usepackage{amsmath,latexsym,amssymb,amsthm}
\usepackage{graphicx}
\usepackage{hyperref}
\usepackage{amsmath}\usepackage{amssymb}
\usepackage{amsthm}

\newcommand{\be}{\begin{equation}}
\newcommand{\ee}{\end{equation}}

\newcommand{\R}{\mathbb{R}}

\renewcommand{\>}{\rangle}
\newcommand{\half}{\tfrac{1}{2}}

\theoremstyle{definition}

\theoremstyle{remark}

\begin{document}

\title{\bf{Is there a classical model of Wigner's friend?}}

\author{Anthony Sudbery$^1$\\[10pt] \small Department of Mathematics,
University of York, \\[-2pt] \small Heslington, York, England YO10 5DD\\
\small $^1$ tony.sudbery@york.ac.uk}

\date{31 October 2021}

\maketitle

\begin{abstract}

``Wigner's friend'' refers to a quantum process of which different observers, following the rules of quantum mechanics, give contradictory descriptions. Lostaglio and Bowles have recently claimed to describe a classical system showing the same effect. It is argued that this claim is not justified. It fails to take account of the different meanings of probability in quantum and classical mechanics.

\end{abstract}

In a famous discussion of the foundations of quantum mechanics \cite{Wigner:consciousness}, Wigner described a situation in which two observers use quantum mechanics to describe the same physical state of affairs, but their descriptions are not compatible. The first observer --- Wigner's friend, who I will call Frieda --- takes a qubit in the state $\tfrac{1}{\sqrt{2}}(|0\> + |1\>)$ and measures it in the basis $\{|0\>,|1\>\}$, observing the result $0$ or $1$; let us suppose it is $0$. Following the directions in her quantum mechanics textbook, she concludes that she and the qubit are together in a state $|00\>$, in an obvious notation. Wigner, however, observing Frieda and the lab from outside, sees that the effect of the measurement is to entangle the qubit with Frieda, and concludes that the resulting state of them both is $|+\> = \tfrac{1}{\sqrt{2}}(|00\> + |11\>)$.
 
These descriptions are not compatible; they have implications that are clearly contradictory. Wigner can confirm his description of the state of the lab by measuring it in the basis $|\pm\> = \tfrac{1}{\sqrt{2}}(|00\> \pm |11\>)$; it follows from his description that the result of the measurement will certainly be $+$. Frieda, however, describes the state as $|00\> = \tfrac{1}{\sqrt{2}}(|+\> + |-\>)$, from which it follows that the result of Wigner's measurement is equally likely to be $|+\>$ or $|-\>$. This statement of probabilities can only be checked by considering an ensemble of systems, i.e. by repeating the experiment in an identical manner. In each repetition the state $|00\>$ can lead to either result, and in a run of $N$ repetitions of the experiment all $2^N$ sequences of results are equally likely; if $N$ is large enough the result ``$-$'' is practically certain to occur. It is a consequence of Wigner's state $|+\>$, however, that the result ``$-$'' will never occur.

This incompatibility can be described formally by considering the logical conjunction of Frieda's state and Wigner's state. This conjunction, the intersection of the two state vectors, is the zero vector, which represents impossibility.

Lostaglio and Bowles \cite{LostaglioBowles} claim that this situation can be replicated in a deterministic classical model. In this model the system consists of a set of four bits $X,X',Y,Y'$ which always have definite values $0$ or $1$, and states of the system are identified with probability distributions on the sample space $\{0,1\}^4$, which can be written as a vector in $\R^4$ which is a convex combination of the sixteen extreme points $(\epsilon_1)_X(\epsilon_2)_{X'}(\epsilon_3)_Y(\epsilon_4)_{Y'}$ where $\epsilon_i \in \{0,1\}$. Such a state is a state of knowledge of an agent; the probabilities (coefficients in the convex combination) measure the strength of the agent's belief in the values of the bits. Lostaglio and Bowles describe a process which leads Wigner and Frieda to assign the following states to the bits. Frieda's state is either
\be
\tfrac{1}{4}\big( 0_X 0_{X'} 0_Y 0_{Y'} + 0_X 0_{X'} 0_Y 1_{Y'} + 0_X 1_{X'} 0_Y 0_{Y'} + 0_X 1_{X'} 0_Y 1_{Y'} \big)
\ee
or
\be
\tfrac{1}{4}\big( 1_X 0_{X'} 1_Y 0_{Y'} + 1_X 0_{X'} 1_Y 1_{Y'} + 1_X 1_{X'} 1_Y 0_{Y'} + 1_X 1_{X'} 1_Y 1_{Y'} \big),
\ee
depending on the result of her measurement. Wigner's state is 
\be
\tfrac{1}{4}\big( 0_X 0_{X'} 0_Y 0_{Y'} + 0_X 1_{X'} 0_Y 1_{Y'} + 1_X 0_{X'} 1_Y 1_{Y'} + 1_X 1_{X'} 1_Y 0_{Y'} \big).
\ee

Unlike the state assignments in the quantum model, there is no contradiction in the above descriptions. The descriptions are different, for the simple reason that Wigner and Frieda have different information, and different areas of ignorance; but in both cases (of Frieda's result) they are compatible, not contradictory. Logically, disjunctions ``P or Q or R or S'' and ``P or S or T or U'' do not contradict each other, and this remains true if the disjunctions have epistemic probabilities attached to them. If Wigner and Frieda were to compare their knowledge, they would both update their states --- to $\half\big( 0_X 0_{X'} 0_Y 0_{Y'} + 0_X 1_{X'} 0_Y 1_{Y'}\big)$ in case (1), to $\half\big( 1_X 0_{X'} 1_Y 1_{Y'} + 1_X 1_{X'} 1_Y 0_{Y'}\big)$ in case (2).

Lostaglio and Bowles now represent a measurement on the 4-bit system by a deterministic process which takes each of the above states to a definite state from which the values of the two bits $X$ and $Y$ (Frieda and her qubit) can be predicted. The result is that Wigner predicts the values $(0,0)$ with probability 1, while Frieda, in both cases, predicts that the values will be either $(0,0)$ or $(1,0)$ with probability  $1/2$ each. Again, these predictions are different, but they are not contradictory. The logical conjunction of the two confirms Wigner's prediction of $(0,0)$.

To see why the situation is different in the two models, classical and quantum, consider, as we did in the quantum case, what would happen if the experiment was repeated a large number of times. Because the four bits have definite values which will be the same in each repetition of the experiment, and the process is deterministic, the result will be the same in every repetition. Frieda's prediction is that in a run of $N$ experiments there are just two possible results: either the result is $(0,0)$ every time, or it is $(1,0)$ every time. Her assignment of probability $1/2$ to the result $(1,0)$ does not imply that that result is bound to occur, contradicting Wigner. In the quantum case, by contrast, the process of measurement is stochastic, the repetitions are independent, and there are $2^N$ possible results; if $N$ is large enough the result $|-\>$, because it has probability $1/2$, is practically certain to occur in one of the repetitions.

The difference arises because the probabilities of quantum mechanics refer to future measurements, and ensembles of such measurements, and these probabilities concern what actually happens stochastically in the real world.  The probabilities in Lostaglio and Bowles's model, on the other hand, are measures of present knowledge, as in thermodynamics, and the measurement is deterministic. As shown above, in the quantum situation the predictions obtained from Wigner's state and Frieda's state do contradict each other, and if they were to compare their states, they would have no way of resolving their disagreement. There is a genuine paradox. This is true whether the nature of the states is epistemic or ontic. The conclusion must be that the theory which generates these predictions --- universally valid quantum mechanics with the projection postulate applied to all measurements --- is untenable. It seems unlikely that any model which might be called ``classical'' could exhibit such contradictions.


\begin{thebibliography}{1}

\bibitem{LostaglioBowles}
Matteo Lostaglio and Joseph Bowles.
\newblock The original {W}igner's friend paradox within a realist toy model.
\newblock {\em Proc. Roy. Soc. A.}, 477:20210273, 2021.
\newblock \href{https://arxiv.org/abs/2102.11032}{arXiv:2102.11032}.

\bibitem{QTmeasurement}
J.~A. Wheeler and W.~H. Zurek, editors.
\newblock {\em Quantum Theory and Measurement}.
\newblock Princeton University Press, 1983.

\bibitem{Wigner:consciousness}
E.~P. Wigner.
\newblock Remarks on the mind-body question.
\newblock In I.~J. Good, editor, {\em The Scientist Speculates}, pages
  284--302. Heinemann, 1962.
\newblock reprinted in \cite{QTmeasurement} pp. 168-181.

\end{thebibliography}

\end{document}